\documentstyle[preprint,aps]{revtex}

\begin{document}

\draft

\title{Massive scalar field in multiply connected flat spacetimes}

\author{Tsunefumi Tanaka\cite{Tan} and William A.\ Hiscock\cite{His}}

\address{Department of Physics, Montana State University, Bozeman,
Montana  59717}

\date{\today}

\maketitle

\begin{abstract}

The vacuum expectation value of the stress-energy tensor $\left\langle
0\left| T_{\mu\nu} \right|0\right\rangle$ is calculated in several multiply
connected flat spacetimes for a massive scalar field with arbitrary
curvature coupling.  We find that a nonzero field mass always decreases the
magnitude of the energy density in chronology-respecting manifolds such as
$R^3 \times S^1$, $R^2 \times T^2$, $R^1 \times T^3$, the M\"{o}bius strip,
and the Klein bottle.  In Grant space, which contains nonchronal regions,
whether $\left\langle 0\left| T_{\mu\nu} \right|0\right\rangle$ diverges on
a chronology horizon or not depends on the field mass.  For a sufficiently
large mass $\left\langle 0\left| T_{\mu\nu} \right|0\right\rangle$ remains
finite, and the metric backreaction caused by a massive quantized field may
not be large enough to significantly change the Grant space geometry.

\end{abstract}

\pacs{ }

\section{INTRODUCTION}
The global topology of spacetime, which is not fixed by the equations of
general relativity, plays an important role in quantum field theory even in
a flat spacetime.  When a spacetime is multiply connected only those modes
of a field that satisfy boundary conditions determined by the topology of
the spacetime are relevant in the calculation of locally measurable
quantities such as the stress-energy tensor $T_{\mu\nu}$.  For example, in
a cylindrical two dimensional spacetime, $R^1({\rm time}) \times S^1({\rm
space})$, the only allowed momentum is an integer multiple of
$\frac{2\pi\hbar}{a}$ where $a$ is the circumference in the closed spatial
direction.  DeWitt, Hart, and Isham \cite{DeHaIs} thoroughly studied the
effects of multiple connectedness of the spacetime manifold (called
M\"{o}biosity), twisting of the field, and orientability of manifold on
$\left\langle 0\left| T_{\mu\nu} \right|0\right\rangle$ for a massless
scalar field in various topological spaces.  In this paper we extend their
work to an untwisted massive scalar field.  We evaluate $\left\langle
0\left| T_{\mu\nu} \right|0\right\rangle$ in four dimensional spacetime
manifolds of the type $R^1({\rm time}) \times\Sigma({\rm space}^3)$ where
$\Sigma$ can be either $S^1 \times R^2$, $T^2 \times R^1$, $T^3$, the
M\"{o}bius strip $M^2 \times R^1$, or the Klein bottle $K^2 \times R^1$.
Both the M\"{o}bius strip and the Klein bottle are examples of
nonorientable manifolds.

Another issue addressed in this paper is the effect of field mass on
chronology protection.  In previous papers
\cite{HiKo,Fro,KiTh,Kli,Bou,Gra,Lyu,TaHi,Lau} it has been shown that
$\left\langle 0\left| T_{\mu\nu} \right|0\right\rangle$ for a massless
scalar field diverges on a chronology horizon in various spacetimes with
closed timelike curves (CTC's).  The backreaction of the metric to this
diverging stress-energy through the Einstein field equations may be able to
prevent formation of CTC's.  However, Boulware's work indicates that
$\left\langle 0\left| T_{\mu\nu} \right|0\right\rangle$ for a massive
scalar field remains finite on the chronology horizon in Gott space
\cite{Bou}.  We have confirmed Boulware's result in Grant space which is
holonomic to Gott space \cite{Lau} and found that $\left\langle 0\left|
T_{\mu\nu} \right|0\right\rangle$ is finite on the chronology horizon
provided the mass of the scalar field is above a lower limit which depends
on the topological identification scale lengths of the spacetime.

In Sec.\ II a general procedure for using point-splitting regularization to
calculate the vacuum expectation value of $T_{\mu\nu}$ for a free massive
scalar field in a flat but multiply connected spacetime is described.  We
then apply this method to spacetimes without CTC's in Sec.\ III and to
those with CTC's in Sec.\ IV.  Throughout our calculations natural units in
which $c = G = \hbar = 1$ are used and the metric signature is $+2$.

\section{CALCULATION OF $\left\langle 0\left| T_{\mu\nu}
\right|0\right\rangle$ IN A MULTIPLY CONNECTED FLAT SPACETIME}
Our calculation of the vacuum expectation value of the stress-energy tensor
is greatly simplified by the facts that all curvature components vanish in
a flat spacetime and that we do not need to deal with renormalization of
the field mass $M$ and curvature coupling constant $\xi$ arising from
interactions.  However, we do have to worry about the topology of the
spacetime manifold since it may allow two points on the manifold to be
connected by multiple geodesics.  In this paper the Minkowski vacuum state
is assumed to be the default vacuum state in all spacetimes considered.
This assumption is defended later by an argument based on a particle
detector carried by a geodesic observer.  Point-splitting regularization
(or the ``method of images'') is used to take multiple connectedness into
account.  The differences in topology of the spacetimes will appear only in
the geodesic distances between image charges and in the number of geodesics
connecting the points.  Once geodesical distances for a particular topology
are found the calculation of $\left\langle 0\left| T_{\mu\nu}
\right|0\right\rangle$ reduces to simple differentiation of the Hadamard
elementary function and taking the coincidence limit.

The stress-energy tensor $T_{\mu\nu}$ is formally defined as a variation of
the action with respect to the metric.  In a flat four dimensional
spacetime the stress-energy tensor for a general free scalar field is given
by
\begin{eqnarray}
	T_{\mu\nu} & = & (1 -2\xi)\phi_{;\mu}\phi_{;\nu} + \left(2\xi -
		\frac{1}{2}\right){\sl g}_{\mu\nu}\phi_{;\alpha}\phi^{;\alpha}
		- 2\xi\phi\phi_{;\mu\nu} \nonumber \\
	&  & {} + 2\xi{\sl g}_{\mu\nu}\phi\Box\phi -
		\frac{1}{2}M^2{\sl g}_{\mu\nu}\phi^2.
	\label{Eq1}
\end{eqnarray}
Note that $\left\langle 0\left| T_{\mu\nu} \right|0\right\rangle$ depends
upon the value of $\xi$ even when the curvature vanishes.  For conformal
coupling $\xi = \frac{1}{6}$; for minimal coupling $\xi = 0$.  We will
allow arbitrary values of $\xi$ to make our results as general as possible.
The scalar field $\phi$ satisfies the Klein-Gordon equation $(\Box_x -
M^2)\phi(x) = 0$.

Since every term in $T_{\mu\nu}$ is quadratic in the field variable
$\phi(x)$, we can split the point $x$ into $x$ and $\tilde{x}$ and take the
coincidence limit as $\tilde{x} \to x$.
\begin{eqnarray}
	T_{\mu\nu} & = & \frac{1}{2}\lim_{\tilde{x} \to x}\left[\left(1 -
		2\xi\right)\nabla_\mu\widetilde{\nabla}_\nu + \left(2\xi -
		\frac{1}{2}\right){\sl
		g}_{\mu\nu}\nabla_\alpha\widetilde{\nabla}^\alpha -
		2\xi\nabla_\mu\nabla_\nu\right. \nonumber \\
	& & \left. {} + 2\xi{\sl g}_{\mu\nu}\nabla_\alpha\nabla^\alpha -
		\frac{1}{2}M^2{\sl g}_{\mu\nu}\right]\left\{\phi(x),
		\phi(\tilde{x})\right\},
	\label{Eq2}
\end{eqnarray}
where $\{A, B\}$ is the anticommutator of $A$ and $B$.  Covariant
derivatives $\nabla_\mu$ and $\widetilde{\nabla}_\nu$ are to be applied
with respect to $x$ and $\tilde{x}$.  We have also symmetrized $T_{\mu\nu}$
over $\phi(x)$ and $\phi(\tilde{x})$.  Before taking the vacuum expectation
value of $T_{\mu\nu}$ we need to define the vacuum state of the spacetimes.
This is nontrivial because some spacetimes lack a global timelike Killing
vector field, which is required to define positive frequency.

The spacetimes we will consider may all be constructed by making
topological identifications of Minkowski space, with coordinates and metric
given by
\begin{equation}
	ds^2 = -(dx^0)^2 + (dx^1)^2 + (dx^2)^2 + (dx^3)^2.
	\label{Eq3}
\end{equation}
In the case of the $R^3 \times S^1$, $R^2 \times T^2$, $R^1 \times T^3$,
the M\"{o}bius strip and the Klein bottle spacetimes, there exists a global
timelike Killing vector field (i.\ e., $\frac{\partial}{\partial x^0}$) and
so does a natural vacuum state.  The effect of multiple connectedness on
the field is only making it periodic in closed spatial directions.  Also,
it allows the field to be twisted, and we will briefly comment of the
effect of twisting and the vacuum energy density in the next section.
Since each section of these spacetimes between the boundaries is identical
to a portion of Minkowski space, we can assume the default vacuum states of
these spacetimes to be identical to that of ordinary Minkowski space.

As an example of a spacetime with no natural vacuum state, we will consider
Grant space in Sec.\ IV.  Although Grant space does have local timelike
Killing vector fields everywhere, they can not be patched together to form
a global timelike Killing vector field \cite{HiKo}.  However, each section
of the spacetime is identical to Minkowski space, and a geodesic observer
will not detect any particles if the spacetime is in the Minkowski vacuum
state.  Hence, we can also use the Minkowski vacuum state for the Grant
space.

The vacuum expectation value of stress-energy tensor is now given by
sandwiching both sides of Eq.\ (\ref{Eq2}) by the Minkowski vacuum state
$|0\rangle$.
\begin{eqnarray}
	\left\langle 0\left| T_{\mu\nu} \right|0\right\rangle & = & \frac{1}{2}
		\lim_{\tilde{x} \to x}\left[\left(1 -
		2\xi\right)\nabla_\mu\widetilde{\nabla}_\nu + \left(2\xi -
		\frac{1}{2}\right){\sl
		g}_{\mu\nu}\nabla_\alpha\widetilde{\nabla}^\alpha -
		2\xi\nabla_\mu\nabla_\nu\right. \nonumber \\
	& & \left. {} + 2\xi{\sl g}_{\mu\nu}\nabla_\alpha\nabla^\alpha -
		\frac{1}{2}M^2{\sl g}_{\mu\nu}\right]G^{(1)}(x, \tilde{x}),
	\label{Eq4}
\end{eqnarray}
where the Hadamard elementary function $G^{(1)}(x, \tilde{x})$ is defined
as
\begin{equation}
	G^{(1)}(x, \tilde{x}) = \left\langle 0\left|\left\{\phi(x),
		\phi(\tilde{x})\right\}\right|0 \right\rangle
	\label{Eq5}
\end{equation}
and satisfies the Klein-Gordon equation $\left(\Box_x -
M^2\right)G^{(1)}(x, \tilde{x}) = 0$.  In Minkowski space
$G^{(1)}(x,\tilde{x})$ is a function of the half squared geodesic distance
$\sigma = \frac{1}{2}{\sl g}_{\alpha\beta}(x^\alpha
-\tilde{x}^\alpha)(x^\beta - \tilde{x}^\beta)$ between two points $x$ and
$\tilde{x}$ and has the form
\begin{eqnarray}
	G^{(1)}(x, \tilde{x}) & = & \frac{M}{2\pi^2
		\sqrt{2\sigma}}\Theta(2\sigma) K_1(M\sqrt{2\sigma})
		\nonumber \\
	&  & {} + \frac{M}{4\pi\sqrt{-2\sigma}}\Theta(-2\sigma)
		I_1(M\sqrt{-2\sigma}),
	\label{Eq6}
\end{eqnarray}
where $\Theta$ is a step function and $I_1$ and $K_1$ are modified Bessel
functions of the first and second kinds, respectively \cite{Ful}.

Because the spacetime is multiply connected, there can be more than one
geodesic connecting the two points $x$ and $\tilde{x}$.  For example,
suppose the spacetime is closed in the $x^1$ direction.  We can connect $x$
and $\tilde{x}$ with a direct path, or we can start from $x$ and circle
around in the $x^1$ direction once, twice, or an arbitrary number of times
before arriving at $\tilde{x}$.  Since the path circling around $n$ times
cannot be deformed continuously into the one which circles around $n'(n'
\not= n$) times, all inequivalent paths must be taken into account.
Equivalently we can consider this situation as an electrostatic problem and
use the method of images.  The ``image charges'' of the point $\tilde{x}$
are located at $\tilde{x}\pm a, \tilde{x}\pm 2a, \cdots, \tilde{x}\pm na$,
where $a$ is the periodicity (or circumference) in the closed spatial
direction.  All these image charges are connected to the point $x$ by
geodesics whose half squared distances $\sigma_n$ are given by
\begin{equation}
	\sigma_n = \frac{1}{2}{\sl g}_{\alpha\beta}(x^\alpha -
	\tilde{x}^\alpha_n)(x^\beta - \tilde{x}^\beta_n),
	\label{Eq7}
\end{equation}
where $\tilde{x}_n$ is the position of the $n$th image charge.  A
contribution from each image charge is summed over to construct the
regularized Hadamard function $G^{(1)}_{\rm reg}$.  However, using
$G^{(1)}_{\rm reg}$ in Eq.\ (\ref{Eq4}) gives infinity because the
stress-energy tensor is not renormalized yet.  The infinite vacuum energy
term associated with the Minkowski vacuum state must be subtracted from
$G^{(1)}_{\rm reg}$.  This term comes from $G^{(1)}$ for the image charge
at $\tilde{x}_0$.  Excluding it from the summation we obtain the
renormalized Hadamard function $G^{(1)}_{\rm ren}$ which, using Eq.\
(\ref{Eq4}), gives the renormalized stress-energy tensor $\left\langle
0\left| T_{\mu\nu} \right|0\right\rangle_{\rm ren}$.
\begin{equation}
	G^{(1)}_{\rm ren}(x, \tilde{x}) = \sum_{n = -\infty \atop n \not=
		0}^{\infty}G^{(1)}(\sigma_n)
	\label{Eq8}
\end{equation}
The calculation of $\left\langle 0\left| T_{\mu\nu} \right|0\right\rangle$
has thus been reduced to (1) writing an appropriate $\sigma_n$ for each
topology, (2) applying the derivative operator in Eq.\ (\ref{Eq4}), and (3)
taking the coincidence limit as $\tilde{x} \to x$.

\section{SPACETIMES WITH A GLOBAL TIMELIKE KILLING VECTOR FIELD}
\subsection{Orientable manifolds}
In this section the vacuum expectation value $\left\langle
0\left|T_{\mu\nu}\right|0\right\rangle$ of the stress-energy tensor of a
massive scalar field is evaluated in four dimensional spacetimes with
$R({\rm time}) \times \Sigma({\rm space}^3)$ topology.  A spacetime of this
type has a global timelike Killing vector field (e.\ g.,
$\frac{\partial}{\partial x^0}$) and thus a natural vacuum state.  The
first topology we investigate is $\Sigma = S^1 \times R^2$.  The manifold
can be either orientable or nonorientable.  In an orientable manifold a
triad obeying the right-hand rule maintains the same handedness as it is
translated around the closed spatial dimension.  In Cartesian coordinates
$(x^0, x^1, x^2, x^3)$ the spacetime is closed in the $x^1$ direction with
periodicity $a$, and the following points are identified:
\begin{equation}
	(x^0, x^1, x^2, x^3) \leftrightarrow (x^0, x^1 + na, x^2, x^3),
	\label{Eq9}
\end{equation}
where $n$ is an integer.  The half squared geodesic distance $\sigma_n$
between the point $x$ and the $n$th image charge at $\tilde{x}_n$ is equal
to
\begin{equation}
	\sigma_n = \frac{1}{2}\left[-(x^0 - \tilde{x}^0)^2 + (x^1 - \tilde{x}^1
		- na)^2 + (x^2 - \tilde{x}^2)^2 + (x^3 - \tilde{x}^3)^2\right].
	\label{Eq10}
\end{equation}
We will be concerned with only a spacelike separation ($\sigma_n > 0$), so
the first term in Eq.\ (\ref{Eq6}) is used for the Hadamard function.  This
is because the intervals between the image charges are always spacelike in
all spacetimes concerned.

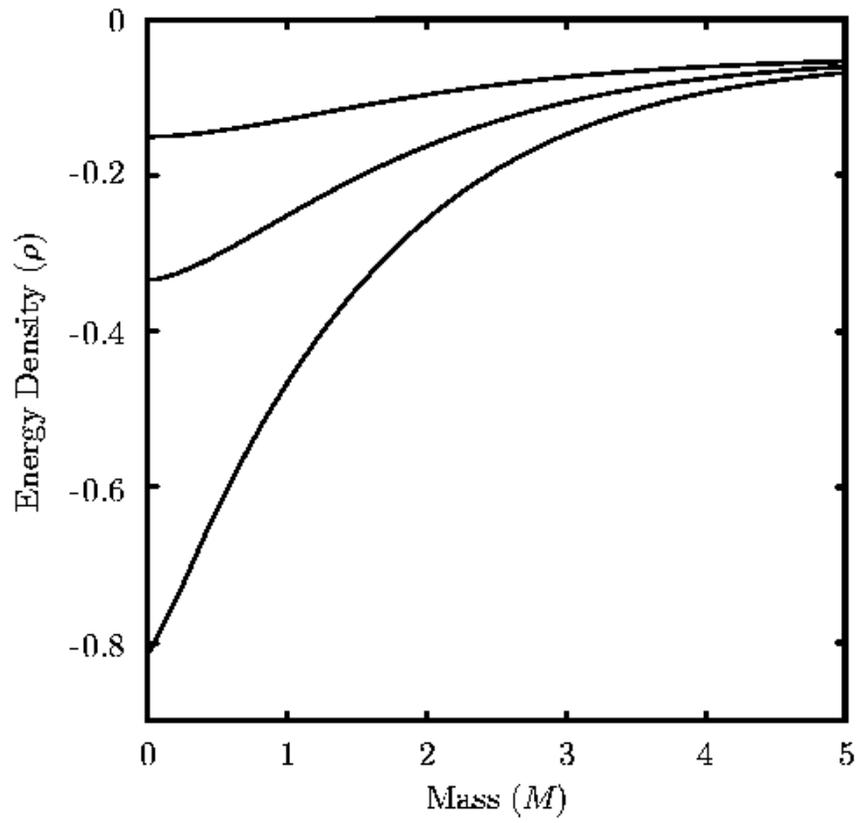
\begin{figure}
	\caption{Plots of the energy density $\rho = \left\langle 0\left| T_{00}
		\right|0\right\rangle$ vs.\ the field mass $M$ for orientable
		manifolds for a conformally coupled field ($\xi = \frac{1}{6}$)
		with periodicities $a = b = c = 1$.  From top to bottom the
		curves represent the topologies $R^3 \times S^1$, $R^2\times
		T^2$, and $R^1 \times T^3$.  Increasing the field mass decreases
		the magnitude of the energy density while increasing
		M\"{o}biosity increases it.}
	\protect\label{Fig1}
\end{figure}

Using the prescription for the calculation of $\left\langle
0\left|T_{\mu\nu}\right|0\right\rangle$ described in the previous section,
we obtain
\begin{equation}
	\left\langle 0\left| T_{\mu\nu} \right|0\right\rangle =
		\frac{M^4}{2\pi^2}\sum_{n =
		1}^{\infty}\left\{\frac{K_2(z_n)}{z_n^2}{\sl g}_{\mu\nu} +
		\frac{K_3(z_n)}{z_n}{\rm diag}[0, -1, 0, 0]\right\},
	\label{Eq11}
\end{equation}
where $z_n = Mna$.  $\left\langle 0\left| T_{\mu\nu} \right|0\right\rangle$
does not depend on the curvature coupling.  The energy density $\rho =
\left\langle 0\left| T_{00} \right|0\right\rangle$ is negative due to a
fact that only certain wavelengths are allowed in the $x^1$ direction
compared to Minkowski space in which all wavelengths are allowed.  In the
massless limit the above result reduces to that given in Ref.\
\cite{DeHaIs}.  Figure \ref{Fig1} shows a plot of $\rho$ vs.  $M$ for $a =
1$.  The effect of mass on the energy density is to raise it (decreasing
its magnitude) because for a larger field mass, its characteristic
wavelength or Compton wavelength $\lambda_C = \frac{1}{M}$ becomes shorter
and is less sensitive to the global structure of the manifold.

The next manifold that we will consider is $\Sigma = T^2 \times R^1$.  The
spacetime is now closed in two spatial dimensions, $x^1$ and $x^2$, with
periodicities $a$ and $b$, respectively.  The following points are
identified
\begin{equation}
	(x^0, x^1, x^2, x^3) \leftrightarrow (x^0, x^1 + na, x^2 + mb, x^3),
	\label{Eq12}
\end{equation}
where $n$ and $m$ are integers.  The half squared geodesic distance
$\sigma_{nm}$ between $x$ and the $nm$th image charge at $\tilde{x}_{nm}$
is given by
\begin{equation}
	\sigma_{nm} = \frac{1}{2}\left[-(x^0 - \tilde{x}^0)^2 +
	(x^1 - \tilde{x}^1 - na)^2 + (x^2 - \tilde{x}^2 - mb)^2 + (x^3 -
	\tilde{x}^3)^2\right].
	\label{Eq13}
\end{equation}
Following the prescription we find
\begin{eqnarray}
	\left\langle 0\left| T_{\mu\nu} \right|0\right\rangle & = &
		\frac{M^4}{4\pi^2}\sum_{n, m = -\infty \atop (n, m) \not=
		(0, 0)}^{\infty}\left\{\frac{K_2(z_{nm})}{z_{nm}^2}
		{\sl g}_{\mu\nu} \right. \nonumber \\
	& & \left. {} + M^2\frac{K_3(z_{nm})}{z_{nm}^3}{\rm diag}[0, -n^2 a^2,
		-m^2 b^2, 0]\right\},
	\label{Eq14}
\end{eqnarray}
where $z_{nm} = M(n^2 a^2 + m^2 b^2)^{\frac{1}{2}}$.  The summation is over
all values of $n$ and $m$ except when $n = m = 0$ simultaneously.  As it
can be seen in Fig.\ \ref{Fig1} $\rho$ for $R^2 \times T^2$ is lower
(greater in magnitude) than that for $R^3 \times T^1$.  The increased
M\"{o}biosity of the manifold lowers the energy density \cite{DeHaIs}.

The last example of orientable spacetime manifold is the one which is
closed in all spatial directions ($\Sigma = T^3$).  The topology of this
spacetime is such that the following points are identified:
\begin{equation}
	(x^0, x^1, x^2, x^3) \leftrightarrow (x^0, x^1 + na, x^2 + mb, x^3
	+ lc),
	\label{Eq15}
\end{equation}
where $n, m$, and $l$ are integers and $a, b$, and $c$ are the
periodicities in the $x^1, x^2$, and $x^3$ directions, respectively.  The
half squared geodesic distance $\sigma_{nml}$ is equal to
\begin{eqnarray}
	\sigma_{nml} & = & \frac{1}{2}\left[-(x^0 - \tilde{x}^0)^2 + (x^1 -
		\tilde{x}^1 - na)^2 + (x^2 - \tilde{x}^2 - mb)^2 \right.
		\nonumber \\
	& & \left. {} + (x^3 - \tilde{x}^3 - lc)^2\right],
	\label{Eq16}
\end{eqnarray}
and the resulting stress-energy tensor is
\begin{eqnarray}
	\left\langle 0\left| T_{\mu\nu} \right|0\right\rangle & = &
		\frac{M^4}{4\pi^2}\sum_{n, m, l = -\infty \atop (n, m, l)
		\not= (0, 0, 0)}^{\infty}\left\{\frac{K_2(z_{nml})}{z_{nml}^2}
		{\sl g}_{\mu\nu} \right.  \nonumber \\
	& & \left. {} + M^2\frac{K_3(z_{nml})}{z_{nml}^3}{\rm diag}[0, -n^2
		a^2, -m^2 b^2, -l^2 c^2]\right\},
	\label{Eq17}
\end{eqnarray}
where $z_{nml} = M(n^2 a^2 + m^2 b^2 + l^2 c^2)^{\frac{1}{2}}$.  The
summation is over all possible value of $n, m$, and $l$ except when $n = m
= l = 0$ at the same time.  As we can see in Fig.\ \ref{Fig1} the increased
M\"{o}biosity lowers the vacuum energy density $\rho$.  Increasing the
field mass raises it toward zero.

The procedure described previous section is readily applicable to the
twisted field.  The only modification required is to introduce the factor
of $(-1)^n$ in the summation in Eq.\ (\ref{Eq8}).  This will make field
antiperiodic.  DeWitt, Hart, and Isham have shown that twisting of the
field raises $\rho$ above the zero-value of Minkowski vacuum state for a
massless scalar field \cite{DeHaIs}.  Our numerical calculations
indicate that this is true even for a massive scalar field.

\subsection{Nonorientable manifolds}
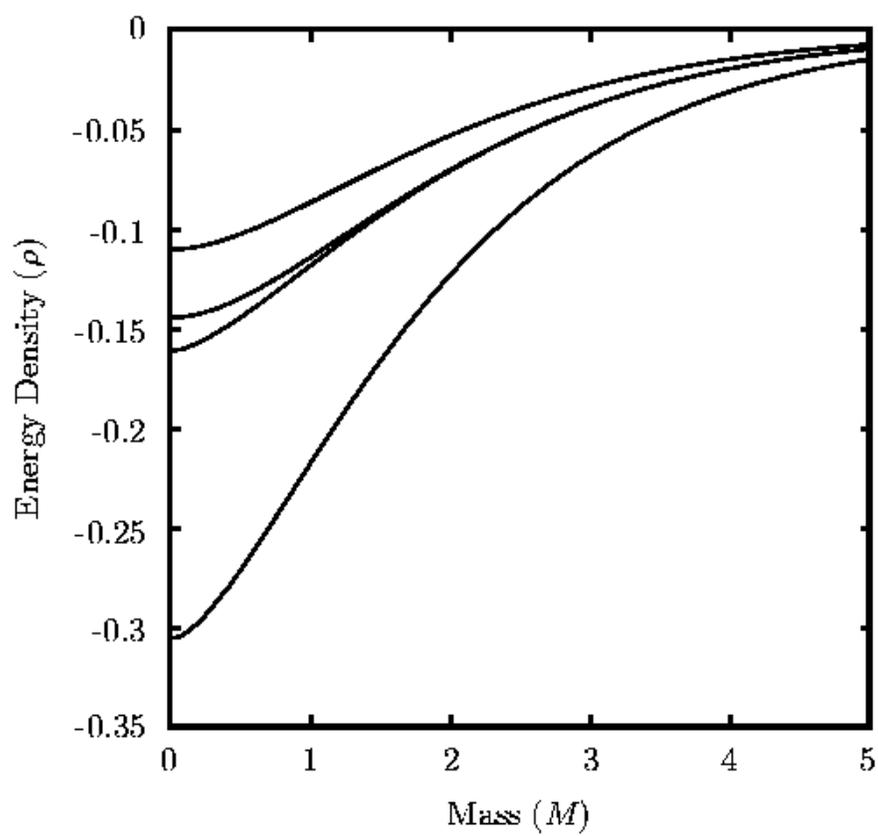
\begin{figure}
	\caption{Plots of the energy density $\rho = \left\langle 0\left| T_{00}
		\right|0\right\rangle$ vs.\ the field mass $M$ for the
		M\"{o}bius strip and Klein bottle at $x^2 = 0$ for a
		conformally coupled field ($\xi = \frac{1}{6}$).  The
		periodicities ($a$ and $b$) are set to 1.
		From top to bottom the curves represent topologies $R^3
		\times S^1$, the M\"{o}bius strip,the Klein bottle, and
		$R^2 \times T^2$.}
	\protect\label{Fig2}
\end{figure}

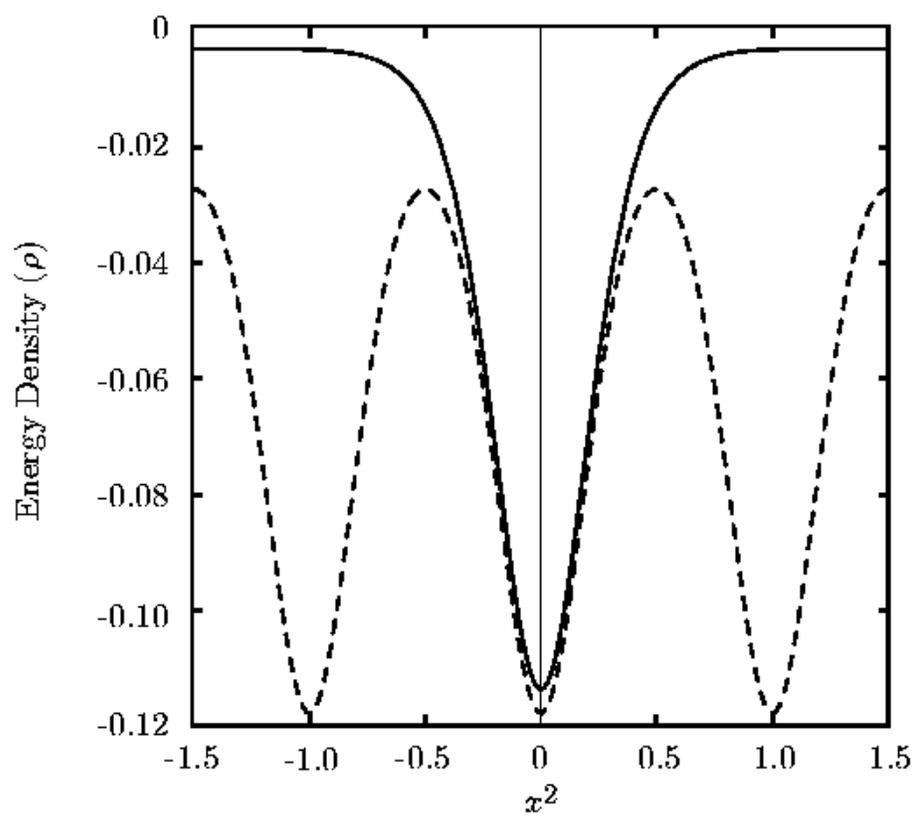
\begin{figure}
	\caption{Plots of the energy density $\rho = \left\langle 0\left| T_{00}
		\right|0\right\rangle$ vs.\ $x^2$ coordinate in the
		M\"{o}bius strip (solid line) and Klein bottle (dashed line)
		with $\xi = \frac{1}{6}$ and $a = b = M = 1$.  $\rho$ decays
		quickly as we move away from $x^2 = 0$ on the M\"{o}bius strip
		whereas it is periodic on the Klein bottle.}
	\protect\label{Fig3}
\end{figure}

The manifolds we have examined so far are
orientable.  Next, we examine how the orientation of the manifold affects
the vacuum polarization.  In a nonorientable manifold a triad which defines
the right-hand rule flips to the one for the left-hand rule when it is
transported around the closed spatial dimension.  The simplest example of a
nonorientable manifold is a M\"{o}bius strip ($M^2$), which locally looks
like the $R^1 \times S^1$ manifold \cite{Wal}.  A four dimensional
M\"{o}bius strip spacetime $R({\rm time}) \times M^2 \times R({\rm space})
$ can be constructed by identifying the points
\begin{equation}
	(x^0, x^1, x^2, x^3) \leftrightarrow (x^0, x^1 + na, (-1)^n x^2, x^3).
	\label{Eq18}
\end{equation}
The half squared geodesic distance $\sigma_n$ is equal to
\begin{equation}
	\sigma_n = \frac{1}{2}\left\{-(x^0 - \tilde{x}^0)^2 + (x^1 - \tilde{x}^1
		- na)^2 + \left[x^2 - (-1)^n \tilde{x}^2\right]^2 + (x^3 -
		\tilde{x}^3)^2\right\}.
	\label{Eq19}
\end{equation}
It is obvious that $\sigma_n$ and thus $\left\langle 0\left| T_{\mu\nu}
\right|0\right\rangle$ will depend on $x^2$ even after the coincidence
limit is taken.
\begin{eqnarray}
	\left\langle 0\left| T_{\mu\nu} \right|0\right\rangle & = &
		\frac{M^4}{2\pi^2}\sum_{n =
		1}^{\infty}\left\{\frac{K_2(z_{2n})}{(z_{2n})^2}{\sl g}_{\mu\nu}
		+ \frac{K_3(z_{2n})}{z_{2n}}{\rm diag}[0, -1, 0, 0]\right\}
		\nonumber \\
	& & {} + \frac{M^4}{4\pi^2}\sum_{n = -\infty}^{\infty}\left\{\left[2(1 -
		2\xi)\frac{K_2(z_{2n + 1})}{(z_{2n + 1})^2} + 4(4\xi -
		1)M^2(x^2)^2\frac{K_3(z_{2n + 1})}{(z_{2n + 1})^3}\right]
		\right.
		\nonumber \\
	& & {} \times {\rm diag}[-1, 1, 0, 1] \nonumber \\
	& & \left. {} + M^2(2n + 1)^2 a^2\frac{K_3(z_{2n + 1})}{(z_{2n +
		1})^3}{\rm diag}[0, -1, 0, 0]\right\},
	\label{Eq20}
\end{eqnarray}
where $z_{2n} = 2Mna$ and $z_{2n + 1} = M[(2n + 1)^2 a^2 +
4(x^2)^2]^{\frac{1}{2}}$.  The first term is identical to that for the case
of $R^3 \times S^1$ with twice the periodicity.  The periodicity is doubled
because the triad must circle around in the $x^1$ direction twice in order
to be oriented in the original sense.  It is interesting that the second
term in Eq.\ (\ref{Eq20}) is dependent on the curvature coupling constant
$\xi$; $\rho$ increases linearly with increasing $\xi$.  For the M\"{o}bius
strip the nonorientability of the manifold lowers the energy density from
that of $R^3 \times S^1$ as we can see in Fig.\ \ref{Fig2}.  Along the
$x^2$ direction $\rho$ has a reverse bell shape centered at $x^2 = 0$.  A
plot of $\rho$ vs.  $x^2$ is shown in Fig.\
\ref{Fig3}.

The last example of a spacetime with $R^1 \times \Sigma$ topology is the
Klein bottle spacetime ($\Sigma = K^2 \times R^1$).  The Klein bottle $K^2$
is a natural extension of the M\"{o}bius strip with an additional spatial
periodic boundary condition.  In Cartesian coordinates, the points to be
identified are
\begin{equation}
	(x^0, x^1, x^2, x^3) \leftrightarrow (x^0, x^1 + na, (-1)^n x^2 + 2mb,
		x^3),
	\label{Eq21}
\end{equation}
and $\sigma_{nm}$ is equal to
\begin{eqnarray}
	\sigma_{nm} & = & \frac{1}{2}\left[-(x^0 - \tilde{x}^0)^2 + (x^1 -
		\tilde{x}^1 - na)^2 + \left\{x^2 - (-1)^n \tilde{x}^2 -
		2mb\right\}^2 \right. \nonumber \\
	& & \left. {} + (x^3 - \tilde{x}^3)^2\right].
	\label{Eq22}
\end{eqnarray}
Again, the resulting $\left\langle 0\left| T_{\mu\nu}
\right|0\right\rangle$ depends on both $x^2$ and $\xi$,
\begin{eqnarray}
	\left\langle 0\left| T_{\mu\nu} \right|0\right\rangle & = &
		\frac{M^4}{4\pi^2}\sum_{n, m = -\infty \atop (n, m) \not=
		(0, 0)}^{\infty}\left\{\frac{K_2(z_{2n, m})}{(z_{2n, m})^2}{\sl
		g}_{\mu\nu} + M^2\frac{K_3(z_{2n, m})}{(z_{2n, m}^3)} \right.
		\nonumber \\
	& & {} \times {\rm diag}[0, -(2n)^2 a^2, -(2m)^2 b^2, 0]\Biggr\}
		\nonumber \\
	& & {} + \frac{M^4}{4\pi^2}\sum_{n, m =
		-\infty}^{\infty}\left\{\left[2(1 - 2\xi)\frac{K_2(z_{2n + 1,
		m})}{(z_{2n + 1, m})^2} \right. \right. \nonumber \\
	& & \left.\left. {} + 4(4\xi - 1)M^2(x^2 -  mb)^2\frac{K_3(z_{2n + 1,
		m})}{(z_{2n + 1, m})^3}\right]{\rm diag}[-1, 1, 0, 1] \right.
		\nonumber \\
	& & \left. {} + M^2 (2n + 1)^2 a^2\frac{K_3(z_{2n + 1, m})}{(z_{2n + 1,
		m})^3}{\rm diag}[0, -1, 0, 0]\right\},
	\label{Eq23}
\end{eqnarray}
where $z_{2n, m} = 2M(n^2 a^2 + m^2 b^2)^{\frac{1}{2}}$ and $z_{2n + 1, m}
= M[(2n + 1)^2 a^2 + 4(x^2 - mb)^2]^{\frac{1}{2}}$.  The first summation is
identical to $\left\langle 0\left| T_{\mu\nu} \right|0\right\rangle$ for
$R^2 \times T^2$ with spatial periodicities doubled in both $x^1$ and $x^2$
directions.  The first summation simply gives a constant negative shift in
the vacuum stress-energy throughout the spacetime.  The second half of
$\left\langle 0\left| T_{\mu\nu} \right|0\right\rangle$ is dependent on
both the $x^2$ coordinate and the curvature coupling $\xi$.  $\rho$ is
oscillatory in the $x^2$ direction with minima at integer multiples of the
periodicity $b$ in that direction (See Fig.\ \ref{Fig3}).  As in the
M\"{o}bius strip spacetime an increase in the value of $\xi$ causes $\rho$
to increase.  However, in contrast to the M\"{o}bius strip spacetime, the
nonorientability of the manifold decreases the magnitude of the energy
density compared to the $R^2 \times T^2$ spacetime as we can see in Fig.\
\ref{Fig2}.

\section{A SPACETIME WITHOUT A GLOBAL TIMELIKE KILLING VECTOR FIELD}
The spacetimes that we have examined so far have one thing in common; they
have a clearly defined Killing time coordinate $x^0$.  In those spacetimes
$\frac{\partial}{\partial x^0}$ is a globally defined timelike Killing
vector field which can be used to define a vacuum state.  However, it is
possible to arrange topological identification of flat space such that a
clearly defined time coordinate for the entire spacetime manifold does not
exist.  As an example of a spacetime without a global timelike Killing
vector field, we will consider Grant space.

Grant space is interesting because it contains closed timelike curves
(CTC's).  The chronology protection conjecture, proposed by Hawking
\cite{Haw}, states that nature prevents the formation of CTC's in spacetimes
which have an initial chronal region.  In order to prove the conjecture,
the protection mechanism within the laws of physics must be identified, if
it exists.  Currently the most promising candidate for the protection
mechanism is the backreaction on the metric due to vacuum polarization of
quantized matter fields.  Quantum field fluctuations pile up on top of
themselves near a chronology horizon causing the vacuum expectation value
of stress-energy tensor $\left\langle 0\left| T_{\mu\nu}
\right|0\right\rangle$ to grow without bound.  This diverging $\left\langle
0\left| T_{\mu\nu} \right|0\right\rangle$ will act back on the metric
through the Einstein field equations and change the spacetime geometry,
possibly preventing the appearance of CTC's.  The vacuum stress-energy of a
massless scalar field has been shown to diverge on the chronology horizon
in Misner space \cite{HiKo}, Gott space \cite{Bou}, wormhole spacetime
\cite{KiTh}, and Roman space \cite{Lyu}.  In our previous paper \cite{TaHi}
we evaluated $\left\langle 0\left| T_{\mu\nu} \right|0\right\rangle$ for a
massive scalar field with arbitrary curvature coupling in Misner space and
found that it diverges on the chronology horizon.  On the other hand,
Boulware \cite{Bou} has shown that the vacuum stress-energy of a massive
scalar field is finite on the chronology horizon of Gott space.  We would
like to know whether that is still true in the Grant space, which is
holonomic to Gott space yet contains Misner space as a special limit.

The original Misner space was developed to illustrate topological
pathologies associated with Taub-NUT (Newman-Unti-Tamburino) type
spacetimes \cite{Mis,HaEl}.  Misner space is simply the flat Kasner
universe with an altered topology.  Its metric in Misner coordinates $(y^0,
y^1, y^2, y^3)$ is
\begin{equation}
	ds^2 = -(dy^0)^2 + (y^0)^2(dy^1)^2 + (dy^2)^2 + (dy^3)^2.
	\label{Eq24}
\end{equation}
That Misner space is flat can be easily seen; the above metric becomes
identical to the Minkowski metric via the following coordinate
transformation:
\begin{equation}
	x^0 = y^0\cosh y^1, \quad x^1 = y^0\sinh y^1, \quad x^2 = y^2,
		\quad x^3 = y^3.
	\label{Eq25}
\end{equation}
Grant space is constructed by making topological identifications in the
$y^1$ and $y^2$ directions:
\begin{equation}
	(y^0, y^1, y^2, y^3) \leftrightarrow (y^0, y^1 + na, y^2 - nb, y^3).
	\label{Eq26}
\end{equation}
Misner space is the special case $b = 0$.  In Cartesian coordinates the
above identification is equivalent to
\begin{eqnarray}
	(x^0, x^1, x^2, x^3) & \leftrightarrow & (x^0\cosh(na) + x^1\sinh(na),
		\nonumber \\
	& & \quad x^0\sinh(na) + x^1\cosh(na), x^2 - nb, x^3).
	\label{Eq27}
\end{eqnarray}
It can be shown that Grant space is actually a portion of (holonomic to)
Gott space, which consists of two infinitely long, straight cosmic strings
passing by each other \cite{Lau,Tho}.  The periodicities $a$ and $b$ in
Grant space are related respectively to the relative speed and distance
between the two cosmic strings in Gott space.  As $b$ approaches zero (the
Misner space limit) the impact parameter of the two strings also approaches
zero.

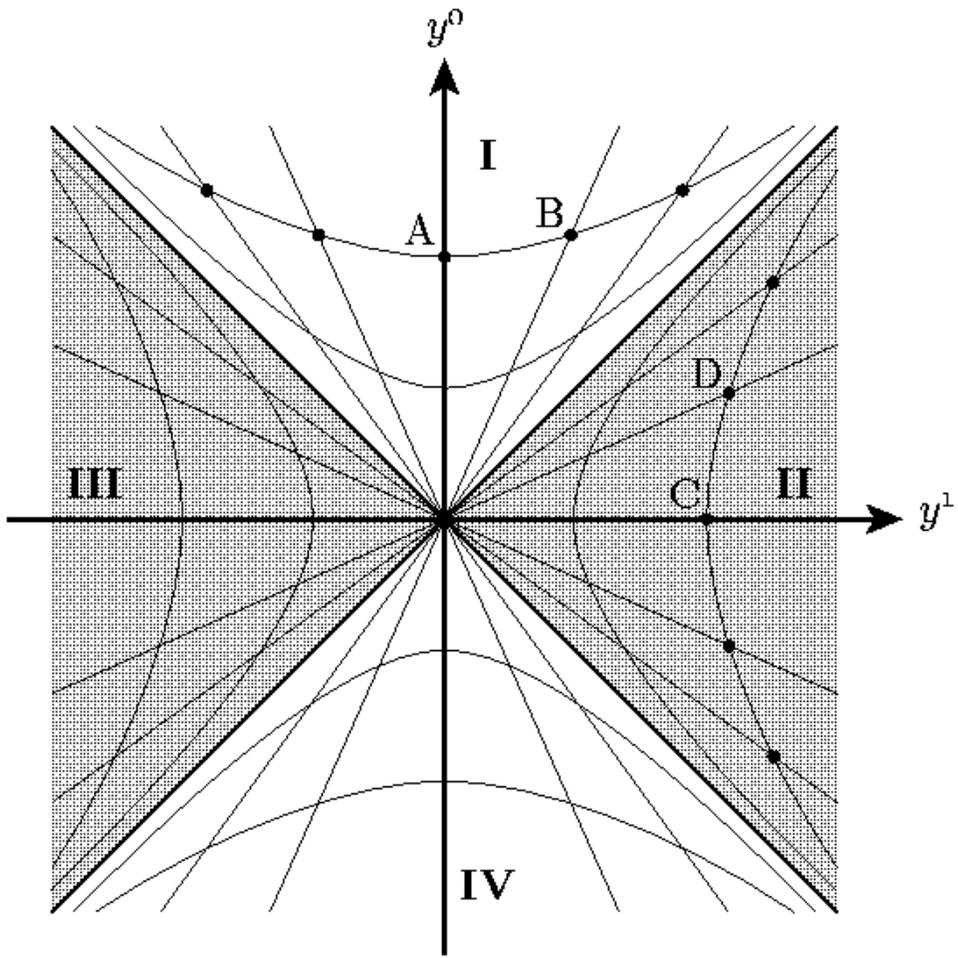
\begin{figure}
\caption{Spacetime diagram of the maximally extended Grant space.  The
		radial straight lines are identified periodic boundaries at
		$y^1 = na$, and the hyperbolas are surfaces of constant $y^0$
		in regions I and IV.  Points A and B are identified with each
		other. As a particle crosses the boundary from one interval to
		next, it is Lorentz boosted in the $x^1$ direction and
		translated in the $x^2$ direction by $-b$.  The roles of $x^0$
		and $x^1$ are switched in nonchronal regions II and III.  Two
		images charges C and D can be connected by a timelike curve.
		CTC's exist in these regions.}
	\protect\label{Fig4}
\end{figure}

Grant space can be considered as a portion of the $R^2 \times T^2$
spacetime of the previous section with the boundaries in the $x^1$
direction moving toward each other at constant velocity.  A spacetime
diagram of the maximally extended Grant space is shown in Fig.\ \ref{Fig4}.
In regions I and IV radial straight lines represent $y^1 = na$ surfaces.
Hyperbolas are constant $y^0$ surfaces.  All image charges of a point are
located on the same hyperbolic surface.  As a particle crosses the radial
boundary, $y^1 = na$, it is Lorentz boosted in a new inertial frame moving
at a speed $v = \tanh a$ in the $x^1$ direction with respect to the
original frame and is translated by $-b$ in the $x^2$ direction.  What is
extraordinary about Grant space is that it contains nonchronal regions (II
and III).  In those regions the roles of $y^0$ and $y^1$ are switched.  The
radial boundaries are now spacelike and the spacetime becomes periodic in
the time ($y^0$) direction.  This topological identification allows the
formation of CTC's in those regions.  The boundaries ($x^0 = \pm x^1$)
separating chronal regions (I and IV) and nonchronal regions (II and III)
are chronology horizons which are a kind of Cauchy horizons.  The
chronological structure of Grant space is discussed in Ref.\ \cite{Tho}.

Within each interval between the periodic boundaries (i.\ e., one period)
$\frac{\partial}{\partial x^0}$ can play the role of a timelike Killing
vector field, but it is impossible to define a global timelike Killing
vector field by patching these $\frac{\partial}{\partial x^0}$'s together.
Without a global timelike Killing vector field the vacuum state of the
spacetime cannot be defined.  However, we argue that the Minkowski vacuum
state is a valid vacuum state of the Grant space.  Each interval in Grant
space is identical to a portion of Minkowski space, so a geodesic observer
in the interval will not detect any particle in the Minkowski vacuum.
Since the only difference between one interval to its neighbors is a
constant relative velocity in the $x^1$ direction and a translation in the
$x^2$ direction, geodesic observers in the neighboring intervals will not
find any particles in the same vacuum state.  The state in which no
geodesic observer detects any particles can be considered as a vacuum
state.  This allows us to use the same renormalized Hadamard function Eq.\
(\ref{Eq6}) in the calculation of $\left\langle 0\left| T_{\mu\nu}
\right|0\right\rangle$ in Grant space.

The result of the calculation is most simply expressed in the Misner
coordinates, ($y^0,y^1,y^2,y^3$),
\begin{eqnarray}
	\left\langle 0\left|T_0^0\right|0\right\rangle & = &
		\frac{M^4}{2\pi^2}\sum_{n = 1}^{\infty}\left[1 +
		4\xi\sinh^2\left(\frac{na}{2}\right)\right]
		\frac{K_2(z_n)}{z_n^2}, \nonumber \\
	\left\langle 0\left|T_1^1\right|0\right\rangle & = &
		\frac{M^4}{2\pi^2}\sum_{n = 1}^{\infty}\left[1 +
		4\xi\sinh^2\left(\frac{na}{2}\right)\right] \nonumber \\
	& & {} \times \left[\frac{K_2(z_n)}{z_n^2} - 4M^2
		(y^0)^2\sinh^2\left(\frac{na}{2}\right)\frac{K_3(z_n)}{z_n^3}
		\right], \nonumber \\
	\left\langle 0\left|T_2^2\right|0\right\rangle & = &
		\frac{M^4}{2\pi^2}\sum_{n = 1}^{\infty}\left\{\left[1 + 2(4\xi -
		1)\sinh^2\left(\frac{na}{2}\right)\right]\frac{K_2(z_n)}{z_n^2}
		\right. \nonumber \\
	& & \left. {} + M^2\left[4(1 - 4\xi)(y^0)^2\sinh^4\left(\frac{na}{2}
	\right) - n^2 b^2\right]\frac{K_3(z_n)}{z_n^3}\right\}, \nonumber \\
	\left\langle 0\left|T_3^3\right|0\right\rangle & = & \left\langle
		0\left|T_{22}\right|0\right\rangle + \frac{M^6 b^2}{2\pi^2}
		\sum_{n = 1}^{\infty}n^2\frac{K_3(z_n)}{z_n^3},
	\label{Eq28}
\end{eqnarray}
where $z_n = M\left[4(y^0)^2\sinh^2\left(\frac{na}{2}\right) + n^2
b^2\right]^{\frac{1}{2}}$.  The trace is equal to
\begin{eqnarray}
	\left\langle 0\left|T_{\mu}^{\mu}\right|0\right\rangle & = &
		\frac{M^4}{2\pi^2}\sum_{n = 1}^{\infty}\left\{4\left[1 + (6\xi
		-1)\sinh^2\left(\frac{na}{2}\right)\right]
		\frac{K_2(z_n)}{z_n^2}
		\right. \nonumber \\
	& & \left. {} + \left[-z_n^2 + 8(1 -
		6\xi)M^2(y^0)^2\sinh^4\left(\frac{na}{2}\right)\right]
		\frac{K_3(z_n)}{z_n^3}\right\}.
	\label{Eq29}
\end{eqnarray}
On the chronology horizon ($y^0 = 0$), the components of the stress-energy
tensor are
\begin{eqnarray}
	\left\langle 0\left|T_0^0\right|0\right\rangle & = &
		\frac{M^4}{2\pi^2}\sum_{n = 1}^{\infty}\left[1 +
		4\xi\sinh^2\left(\frac{na}{2}\right)\right]
		\frac{K_2(Mnb)}{(Mnb)^2}, \nonumber \\
	\left\langle 0\left|T_1^1\right|0\right\rangle & = & \left\langle
		0\left|T_0^0\right|0\right\rangle, \nonumber \\
	\left\langle 0\left|T_2^2\right|0\right\rangle & = &
	\frac{M^4}{2\pi^2}\sum_{n = 1}^{\infty}\left\{\left[1 + 2(4\xi -
		1)\sinh^2\left(\frac{na}{2}\right)\right]
		\frac{K_2(Mnb)}{(Mnb)^2} - \frac{K_3(Mnb)}{Mnb}\right\},
		\nonumber \\
	\left\langle 0\left|T_3^3\right|0\right\rangle & = &
		\frac{M^4}{2\pi^2}\sum_{n = 1}^{\infty}\left[1 + 2(4\xi -
		1)\sinh^2\left(\frac{na}{2}\right)\right]
		\frac{K_2(Mnb)}{(Mnb)^2}.
	\label{Eq30}
\end{eqnarray}

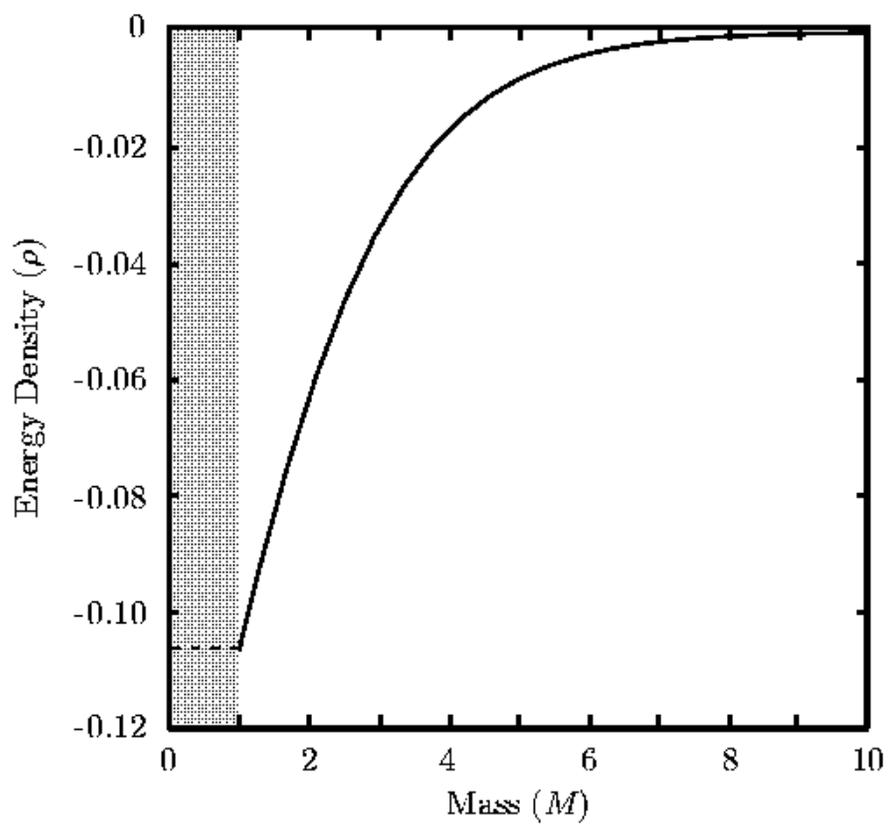
\begin{figure}
	\caption{The energy density of a massive conformal scalar field on the
		chronology horizon in Grant space.  The periodicities $a$ and
		$b$ are both set to 1.  In the shaded region ($M <
		\frac{a}{b} = 1$) $\rho$ diverges and possibly prevents
		formation of CTC's by back reaction on spacetime geometry.
		For $M > \frac{a}{b}$, $\rho$ remains finite on the chronology
		horizon.  At the critical value $M = \frac{a}{b} = 1$, the
		limiting value of $\rho$ is equal to $-0.106$.}
	\protect\label{Fig5}
\end{figure}

Figure \ref{Fig5} shows how the energy density $\rho$ depends on the field
mass $M$ for the conformal coupling ($\xi = \frac{1}{6}$).  For $M <
\frac{a}{b}$, $\rho$ diverges on the chronology horizon.  The contribution
to $\rho$ from the $n$th image charge for $n \gg 1$ is proportional to
$\exp[n(a - Mb)]$.  The factor of $e^{na}$ comes from the Doppler shift as
the particle is boosted in the $y^1$ direction $n$ times.  In Misner space
this factor causes $\left\langle 0\left| T_{\mu\nu} \right|0\right\rangle$
to diverge on the chronology horizon, and it might prevent the formation of
CTC's.  However, in Grant space the exponentially decaying factor
$e^{-nMb}$, which comes from the nonvanishing geodesical distances between
image charges in the $y^2$ direction $b$, makes $\left\langle 0\left|
T_{\mu\nu} \right|0\right\rangle$ finite for values of $M > \frac{a}{b}$.
This result agrees with Boulware's similar calculation in Gott space
\cite{Bou}.  This result implies that the metric backreaction may not be
large enough to significantly change the Grant space geometry if the field
mass is sufficiently large.

\section{DISCUSSION}
In calculating the vacuum expectation value of the stress-energy tensor for
a massive scalar field with arbitrary curvature coupling in multiply
connected flat spacetimes, we have shown that a nonzero field mass raises
the value of the energy density (decreasing its magnitude) and confirmed
that increased M\"{o}biosity lowers the energy density.  The introduction
of nonorientability can evidently either increase the magnitude of the
energy density (M\"{o}bius band) or decrease it (Klein bottle).

Another effect on the stress-energy that we have not examined yet is
self-interaction (e.\ g., $\lambda\phi^4$ theory).  The calculation of
$\left\langle 0\left| T_{\mu\nu} \right|0\right\rangle$ for a
self-interacting field is much more complicated than that for a free field
since it requires renormalization of quantities such as the field mass,
which will be dependent on the field $\phi$, the curvature coupling
constant $\xi$, and the field coupling constant $\lambda$.  In spacetimes
with relatively simple topology, such as $R^3 \times T^1$ and Casimir-type
spacetime, $\left\langle 0\left| T_{\mu\nu} \right|0\right\rangle$ has been
calculated by Birrell and Ford, Ford, Ford and Yoshimura, Kay, and Toms
\cite{For,FoYo,Kay,BiFo,Tom1,Tom2}.  In a spacetime with CTC's
self-interaction is known to cause failure of unitarity
\cite{FrMoNoEcKlThYu,EcKlTh,Nov,FrPaSi1,FrPaSi2}.

In Grant space we discovered that the divergence of $\left\langle 0\left|
T_{\mu\nu} \right|0\right\rangle$ on the chronology horizon depends on
relative size of the field mass to the ratio of the periodicities.  This
result may have significant consequences for chronology protection.  It
suggests that the metric backreaction from the stress-energy of a massive
quantized field will likely not be large enough to significantly alter the
geometry and prevent the formation of CTC's.  If quantized matter fields
are to provide the chronology protection mechanism, our result would
indicate that only massless fields may be capable of providing a
sufficiently strong backreaction to prevent the formation of CTC's.
Outside the domain of quantum gravity, this would place a heavy
responsibility on the electromagnetic field (and conceivably neutrino
fields, should any be massless) as the sole protector of chronology.

\acknowledgements
This work was supported in part by National Science Foundation Grant No.
PHY92-07903.

\end{document}